\documentclass[english,notitlepage,twocolumn,aps,prl,10pt,floatfix,nofootinbib,twocolumn,superscriptaddress]{revtex4-1}
\usepackage[T1]{fontenc}
\usepackage[latin9]{inputenc}
\usepackage{amsmath}
\usepackage{graphicx}
\usepackage{color}
\usepackage{comment}
\usepackage{enumitem,kantlipsum}
\usepackage[normalem]{ulem}
\makeatletter
\usepackage{babel}

\makeatother
\begin{document}
 
\title{Primordial black holes from fifth forces} 
\author{Luca Amendola}
\email{l.amendola@thphys.uni-heidelberg.de}
\author{Javier Rubio}
\email{j.rubio@thphys.uni-heidelberg.de}
\author{Christof Wetterich}
\email{c.wetterich@thphys.uni-heidelberg.de}
\affiliation{Institut f\"ur Theoretische Physik, Ruprecht-Karls-Universit\"at Heidelberg, \\
Philosophenweg 16, 69120 Heidelberg, Germany}
\begin{abstract} 
Primordial black holes can be produced by a long range attractive fifth force stronger than gravity, mediated by a light scalar field 
interacting with nonrelativistic ``heavy" particles. As soon as the energy fraction of heavy particles reaches a threshold, 
the fluctuations rapidly become nonlinear. The overdensities collapse into black holes or similar screened objects, 
without the need for any particular feature in the spectrum of primordial density fluctuations generated during 
inflation. We discuss whether such primordial black holes can constitute the total dark matter component in the Universe.

\end{abstract}

\maketitle
\section{Introduction} 
In spite of the many observations leading to the establishment of dark matter as an 
essential ingredient of modern cosmology, its fundamental nature remains an open question.  Among
the many dark matter candidates, primordial 
black holes (BH) \cite{Zeldovich1966,Hawking:1971ei,Carr:1975qj,Chapline1975} are 
interesting since they could account for the gravitational wave signals observed by the Laser Interferometer 
Gravitational Wave Observatory (LIGO) and the 
VIRGO observatory \cite{ref:LIGO} or seed the formation of supermassive black holes \cite{Kawasaki:2012kn} (see also \cite{Carr:2016drx} and references therein).  The existence of primordial BHs could be a natural consequence of inflation. In particular, if the inflationary potential contains a nontrivial feature along the inflaton trajectory, the 
spectrum of primordial perturbations might develop a peak at intermediate scales. If the amplitude of this
peak is large enough, the nonlinear perturbations will collapse into BHs after horizon reentry \cite{ref:inflection}. 
Alternatively, the formation of primordial BHs could take place at phase transitions  \cite{ref:bubble} or be 
associated with the fragmentation of a scalar condensate into Q-balls \cite{ref:Qballs}. 

In this paper we present a novel framework for primordial BH formation which does not rely on a
particular feature in the spectrum of primordial density fluctuations generated during inflation. The main assumption of our scenario 
is the presence in the early Universe of a long-range interaction stronger than gravity. We associate this 
\textit{fifth force} to a light scalar field  interacting with some heavy degrees 
of freedom beyond the Standard Model particle content. More precisely, we assume that during some epoch 
in cosmology the Hubble parameter $H$ is larger than the mass of a scalar field $\phi$. If this scalar
field couples to some ``heavy particles'' $\psi$ with masses larger than $H$, it mediates an attractive fifth force 
which is effectively long range, similar to gravity. This attraction can be, however, substantially stronger 
than the gravitational attraction. As a result, the fluctuations in the energy density of the heavy fields can grow 
rapidly and eventually become nonlinear. If the range and strength of the fifth force
is large enough, it seems likely that a substantial part of the $\psi$ fluid will collapse into BHs or similar screened objects. 

The existence of long-range attractive forces stronger than gravity is a natural expectation in particle physics 
models containing scalar fields. The simplest example is the attractive interaction among Standard Model
fermions via Higgs particle exchange. In early cosmology it is much stronger than the gravitational attraction. If
not countered by electromagnetic interactions, even the Standard Model Higgs would have induced gravitational collapse
at early times. Alternatively, the $\phi$ and $\psi$ fields could be associated with grand unified frameworks involving,
for instance, a scalar triplet interacting with heavy neutrinos \cite{Magg:1980ut}. In this case, the BH formation 
process could occur 
very early in cosmology, for example nearly after the end of inflation.  For different properties of the participating 
particles it could also take  place rather late in the cosmological history, say after nucleosynthesis. The 
heavy particles remaining outside primordial BHs might decay after the formation epoch and be unobservable today.  The scalar field could relax 
after BH formation to a minimum of its effective potential  with mass eventually exceeding the decreasing Hubble parameter. In this case the field $\phi$ would not be 
observable at the present time either. Alternatively, $\phi$
 could be an additional dark matter candidate, or have a runaway behavior and be associated with dynamical dark energy. 
 The BH formation process is not affected by 
 what happens to the participating fields or particles at later times. Once BHs are formed, they behave as 
 nonrelativistic matter. If the total energy density of BHs is large enough, they could constitute the dark 
 matter component of our Universe.

\section{Fifth-force interactions}
\noindent Consider the action 
\begin{equation}\label{eq:action}
S=\int d^4x \sqrt{-g}\left[\frac{M_{P}^{2}}{2}R+{\cal L}_{R}+{\cal L(\phi)}+{\cal L} (\phi,\psi)\right]\,,
\end{equation} 
with $M_{P}=(8\pi G)^{-1/2}=2.435\times 10^{18} \,{\rm GeV}$ the reduced Planck 
mass and ${\cal L}_R$ a radiation component that we assume to dominate the background evolution during 
the epoch relevant for primordial BH formation. The term
\begin{equation}\label{eq:action2}
{\cal L(\phi)}=-\frac{1}{2}\partial^{\mu}\phi\partial_{\mu}\phi-V(\phi)\,,
\end{equation}
stands for the 
Lagrangian density of an almost massless scalar field $\phi$. We will assume for simplicity that the potential 
$V(\phi)$ can be neglected during the epoch of BH formation such that $\phi$ becomes effectively a massless field. 
This approximation is justified if the scalar field mass is smaller than $H$, both for a potential with 
a minimum or for a runaway potential. For definiteness we take the heavy particle $\psi$ to
be a fermion. The interactions  with the field $\phi$ arise via a field-dependent mass term $m(\phi)$,
\begin{equation}\label{eq:action3}
{\cal L(\phi,\psi)}=i\bar{\psi}\left(\gamma^{\mu}\nabla_{\mu}-m(\phi)\right)\psi\,,
\end{equation}
for example by a Yukawa type coupling $\sim g\phi \bar\psi\psi$. 

We will assume that the field equations derived from the action \eqref{eq:action} admit a perfect 
fluid description and consider a flat Friedmann-Lema\^itre-Robertson-Walker Universe. 
The background evolution equations for the average $\phi$ and $\psi$ energy densities read
\begin{eqnarray}
 && \dot{\rho}_{\phi}+3H(\rho_{\phi}+p_{\phi})=\,\frac{\beta}{M_{P}}\left(\rho_{{\rm \psi}}-3p_\psi\right)\dot{\phi}\,,\label{eq3a}\\
 && \dot{\rho}_{{\rm \psi}}+3H\left(\rho_{{\rm \psi}}+p_\psi\right)=-\,\,\frac{\beta}{M_{P}}\left(\rho_{{\rm \psi}}-3p_\psi\right)\dot{\phi}\,,\label{eq3}
\end{eqnarray}
with $\rho_\phi=p_\phi=\dot\phi^2/2$, $H^2=\rho/(3M_P^2)$ and $\rho=\rho_R+\rho_\phi+\rho_\psi$. The $\phi$ and $\psi$ fluids are 
coupled whenever the $\psi$ particles are nonrelativistic ($\rho_\psi\neq 3 p_\psi$, with $p_\psi$ the $\psi$-fluid pressure). The 
coupling function $\beta(\phi)$ 
measures the dependence of the effective mass $m(\phi)$ on the scalar field $\phi$, 
\begin{equation}
\beta(\phi)=-M_{P}\frac{\partial\ln m(\phi)}{\partial\phi}\,.
\label{betadef1}
\end{equation}
Its normalization involving $M_P$ has been chosen such that for $\beta^2=1/2$ the scalar-field mediated attraction 
 has the same strength as gravity. The combined strength of the fifth force and gravity 
is proportional to 
\begin{equation}\label{eq:Ydef}
Y\equiv 1+2\beta^2\,. 
\end{equation}

This type of scenario has been extensively studied in the literature \cite{Wetterich:1994bg,Fardon:2003eh,Brookfield:2005bz,Wetterich:2007kr,Amendola:2007yx,Casas:2016duf}, 
 but not within the context of primordial black hole formation.  A given model is specified by a choice of $\beta(\phi)$. The value of $\beta$ can
 be rather large.  Consider, for instance, a renormalizable interaction
 term of the form $m(\phi)\bar\psi\psi=m_0\bar\psi\psi+g\phi\bar\psi \psi$
with $m_0$ a constant mass parameter and $g$ a dimensionless coupling. For this particular example, we can 
 rewrite Eq.~\eqref{betadef1} as $\beta(\phi)=-g\,M_P/m(\phi)$, 
 which leads to $\vert \beta\vert \gg 1$ if $m(\phi)/M_{P}\ll g$. Even a small value of the Yukawa 
coupling $g$ can be largely overwhelmed by the ratio $M_P/m$. The factor $2\beta^2$  
can therefore be naturally rather large. 
For illustration purposes we will neglect the field dependence of $\beta$ in the following 
sections. The qualitative features of the BH formation scenario presented here do not 
rely on this approximation. They also hold if we take for $\psi$ a nonrelativistic bosonic particle rather than a 
fermion.

\section{Background cosmology}
For a nonrelativistic fluid of $\psi$ particles, $p_\psi=0$, the set of equations \eqref{eq3a}-\eqref{eq3} 
admits a scaling solution. In the limit $\beta\gg1$ it reads 
\cite{Amendola:1999er}
\begin{equation}
\Omega_{\psi}=\frac{1}{3\beta^{2}}\,,\hspace{8mm}
\Omega_{\phi}=\frac{1}{6\beta^{2}}\,,\hspace{8mm}
  \Omega_{R}=1-\frac{1}{2\beta^{2}}\,,
\label{eq:phaseq-1}
\end{equation}
where $\Omega_{i}\equiv \rho_{i}/(3M_{P}^{2}H^{2})$ and $i=R,\phi,\psi$.
During this scaling phase, the fermion energy density \textit{tracks} the background component, 
$\rho_{\psi}\sim\rho_{R}\sim a^{-4}$. Combining this scaling with the intuitive solution of Eq.~\eqref{eq3}, $
\rho_\psi=\rho_{\psi,0}\, a^{-3} \exp\left(-\beta\phi/M_P\right)$, 
we obtain a relation between $\beta$ and the variation of the $\phi$ field during the scaling phase, namely
\begin{equation}\label{eq:bphi}
\phi'=M_P/\beta\,,
\end{equation}
with primes denoting derivatives with respect to the number of $e$-folds $dN\equiv d\ln a$, e.g. $\phi'=\dot\phi/H$.
\begin{figure}
\begin{center}
\includegraphics[scale=0.27]{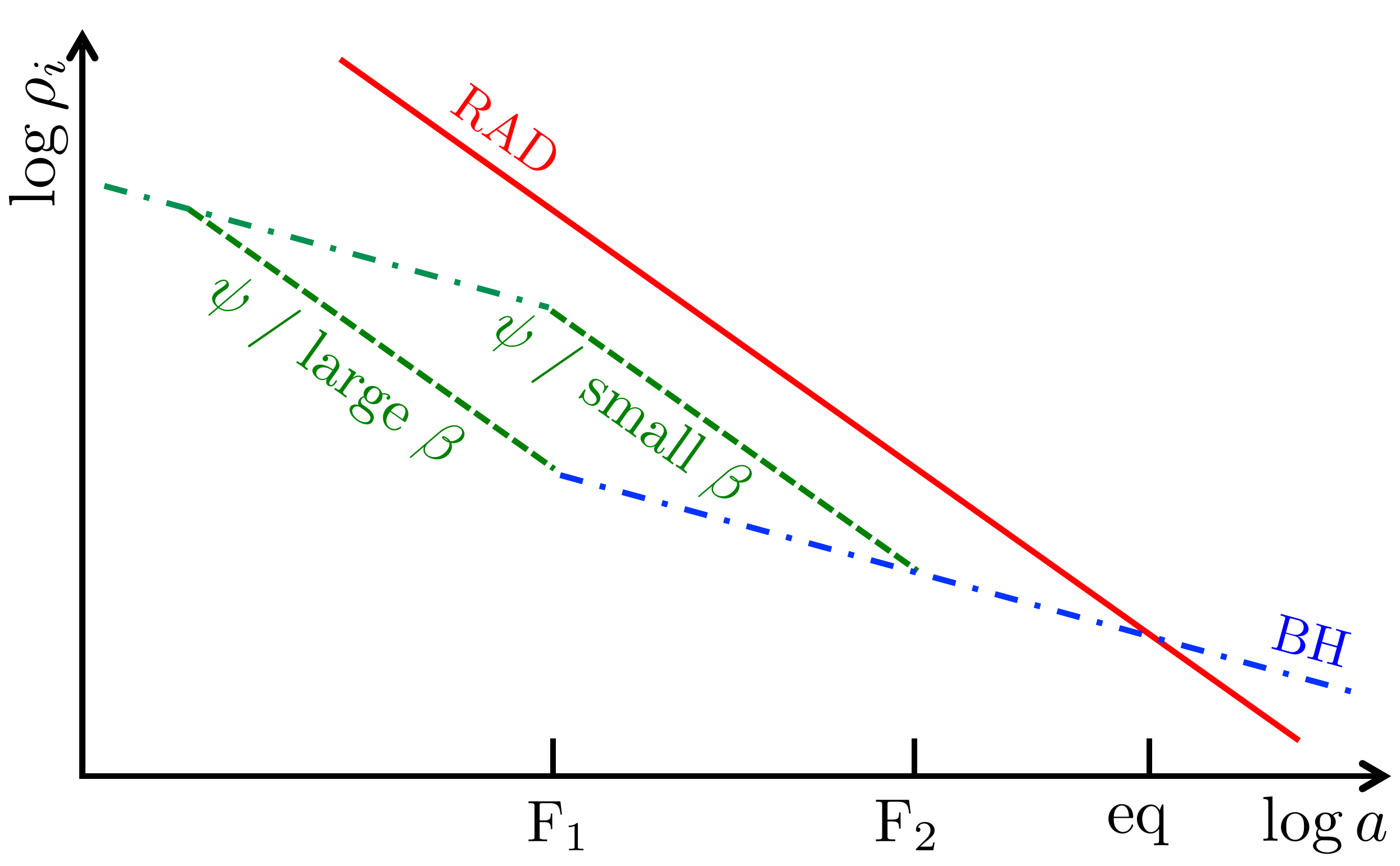}
\caption{Schematic evolution of the energy density of different species. The red and  blue lines stand respectively 
for radiation and BH dark matter. Green lines denote $\rho_\psi$. Larger values of $\beta$ imply earlier BH formation and therefore 
smaller masses. For $\Omega_{\rm BH}(a_{\rm eq})<1/2$ the radiation line moves upwards.}\label{fig:evol}
\end{center}
\end{figure}

Alternatively, $\Omega_\psi$ could be even smaller than $1/(3\beta^2)$. The fifth force then plays
no role for the background evolution and $\rho_\psi$ 
decays as nonrelativistic matter, $\rho_\psi\sim a^{-3}$. In consequence, the  density parameter $\Omega_\psi$ increases with
time until it  reaches the scaling solution $\Omega_\psi=1/(3\beta^2)$ at some time $t_{\rm in}$.  The evolution of the 
different energy densities is depicted in Fig.~\ref{fig:evol}.

Which of the above regimes is realized during the BH formation period depends on the initial conditions for 
the energy densities $\rho_\psi$ and $\rho_\phi$, which are typically set at the end of inflation.
For the large $\beta^2$ values we will be interested in, both the difference between various initial 
conditions and the question of whether or not the scaling solution \eqref{eq:phaseq-1} is reached become
 unimportant from the point of view of the background evolution. For all practical purposes, the background behaves 
 as a standard radiation-dominated (RD) Universe with a Hubble parameter obeying ${{\cal H}'}=-{{\cal H}}$, with ${\cal H}=aH$.

\section{Growth of fluctuations}
Due to the strong  attractive force for $Y\gg 1$ and the decreasing particle masses, the 
$\psi$ fluctuations grow rapidly for sufficiently large $\Omega_\psi$. On scales $k^{-1}$ sufficiently inside the horizon,  the linear 
perturbations  $\delta_\psi$ in the $\psi$ fluid evolve as \cite{Amendola:2003wa}
\begin{eqnarray}\label{pertpsi-1}
\hspace{-6mm}\delta_{\psi}''&+& \left(\hspace{-0.5mm}1 \hspace{-0.5mm}+\hspace{-0.5mm}\frac{\mathcal{H}'}{\mathcal{H}}
-\frac{\beta\phi'}{M_P}\right)\delta'_{\psi}-\frac{3}{2}
(Y\Omega_{\psi}\delta_{\psi}+\Omega_{R}\delta_{R})  =0.  
\end{eqnarray} 
For large $\beta^2$ the large value of $Y$ 
accounts for the enhanced attraction between the $\psi$ particles. In addition we observe a generalized 
damping (or rather antidamping) term $\sim \beta\phi'$ due to the change of the heavy particle mass as $\phi$ evolves. 
 The perturbations in $\rho_\phi$ are negligible in view of a unit sound speed. The perturbations $\delta_R$ in the radiation 
 fluid follow the standard behavior and do not grow. Their small amplitude remains at the level 
inherited from inflation. 

Even for a very small initial perturbation $\delta_\psi$ the inhomogeneities in the
radiation fluid will trigger inhomogeneities in the $\psi$ fluid due to the source term $\delta_R$ in the 
evolution equation for $\delta_\psi$.  This typically implies a minimal value for $\delta_\psi$ of the same 
order as $\delta_R$. For the further growth of $\delta_\psi$ we can neglect $\delta_R$.

We first discuss the growth rate for the scaling solution \eqref{eq:phaseq-1}. Inserting Eq.~\eqref{eq:bphi}, together with $Y\Omega_\psi\simeq 2/3$ and ${{\cal H}'}=-{{\cal H}}$, and neglecting $\delta_R$, 
the evolution equation \eqref{pertpsi-1} becomes rather simple,
\begin{align}\label{eq:growthscaling}
\delta_{\psi}''-\delta'_{\psi}-\delta_{\psi} & =0\,.
\end{align}
The solution of this differential equation contains a growing and a decaying mode. At large 
$a$, we are left only with the growing piece, namely
\begin{equation}\label{eq:pertgrowth}
\delta_\psi=\delta_{\psi, \rm in}\left(a/a_{\rm in}\right)^{p}\,,\hspace{8mm} p=(1+\sqrt{5})/2\approx 1.62\,,
\end{equation}
with $a_{\rm in}\equiv a(t_{\rm in})$ the scale 
factor at the onset of the scaling regime. The $\phi$-$\psi$ interactions in Eqs. \eqref{eq3a} and \eqref{eq3}
translate into a power-law growth of the $\psi$ perturbations during the scaling regime.  
This is a key difference with respect to nonrelativistic matter without the fifth force.  For $\beta=0$ the fluctuations 
$\delta_\psi$ do not grow during radiation domination. 

The scaling solution is not essential for the growth of $\psi$ perturbations. 
For $\Omega_\psi>1/(3\beta^2)$ the growth is faster than for the scaling 
solution, while for $\Omega_\psi<1/(3\beta^2)$ the growth slows down. The growth also slows if $\phi$ changes 
slower than the scaling solution \eqref{eq:bphi}. Furthermore, the growth rate decreases as the wavelength of fluctuations 
increases towards the horizon. Many scenarios can be covered by treating $p$ 
as approximately constant and roughly of the order one.  For constant $p$ even small initial 
inhomogeneities $\delta_\psi$, e.g. $\delta_{\psi,{\rm in}}\sim 10^{-5}$, will develop into nonlinear 
inhomogeneities rather rapidly. The number of $e$-folds  for the onset of nonlinearity is 
\begin{equation}\label{eq:onset_nl}
N_{\rm F}\equiv \ln\left(\frac{a_{\rm F}}{a_{\rm in}}\right)=\frac{1}{p}\ln\left(\frac{\delta_c}{\delta_{\psi,{\rm in}}}\right) \,.
\end{equation}
with $\delta_\psi(a_{\rm F})\equiv \delta_c\sim {\cal O}(1)$. This number of $e$-folds is typically much smaller 
than the duration of the RD era. For $\delta_{\psi,{\rm in}}$ comparable to 
$\delta_R\approx 10^{-5}$ and $p=(1+\sqrt{5})/2$, it only takes $N_{\rm F}\approx 7$ $e$-folds before the 
fluctuations become nonlinear.

We conclude that the fluctuations in the energy density of a nonrelativistic fluid always 
become nonlinear if the following conditions are satisfied: (i) the coupling $\beta$ is large, (ii)  
the fraction $\Omega_\psi$ reaches a value of the order $\beta^{-2}$ early enough in the RD era, 
such that a value $p\approx 1$ is realized, and (iii) the scalar field $\phi$ has a mass smaller than $H$ 
during the growth period. 

An initial $\Omega_\psi$ much smaller than $1/(3\beta^2)$ grows until it reaches values of the order $1/(3\beta^2)$ at 
$t_{\rm in}$. During this epoch $p$ is small and the 
growth of fluctuations remains moderate. Once $\Omega_\psi$ reaches the scaling regime, 
$\Omega_\psi\approx 1/(3\beta^2)$, the   scaling solution with a fast growth of 
perturbations \eqref{eq:pertgrowth} becomes a good approximation. For small initial 
$\Omega_\psi$ this simply sets the time $t_{\rm in}$ for the onset of the growth to the time 
at which $\Omega_\psi$ reaches the value $\approx 1/(3\beta^2)$. For more detailed considerations 
one may employ in Eq.~\eqref{eq:pertgrowth} a growth rate $p(a)$ that depends on $a$. 

\section{Black hole formation}
When the fluctuations become nonlinear one expects the collapse of overdense regions. A collapsing overdense region 
of the size of the horizon will presumably form a black hole if nothing stops the approximately spherical infall. An alternative 
would be the formation of highly concentrated lumps. Black holes 
interact only by gravitational forces. Thus the $\psi$-particles caught 
in black holes no longer feel the fifth force. A similar screened behavior can be expected for highly concentrated lumps due 
to strong backreaction effects \cite{Ayaita:2012xm,Ayaita:2011ay}. For cosmological purposes black holes and screened objects behave similarly and we 
will no longer make the distinction.

The BH formation subtracts from the energy density $\rho_\psi$ a fraction 
$\rho_{\rm BH}$ that is converted into BHs. Black holes remain stable even if the $\psi$ particles outside them 
decay at a later stage. The BH fluid is a nonrelativistic fluid contributing to dark matter, with $\rho_{\rm BH}\sim a^{-3}$ once 
no new BHs are generated, cf. Fig.~\ref{fig:evol}. A rather small density parameter $\Omega_{\rm BH}$ generated early in cosmology  will grow $\sim a$ during the long RD period and can reach substantial values. During the BH formation period 
$\Omega_{\rm BH}$ will grow even faster than $\sim a$. We neglect here accretion effects after BH formation which 
enhance the growth of $\rho_{\rm BH}$.

There is a simple relation between the time when a black hole forms, as expressed by the
scale factor $a_{\rm F}$ at collapse, and the mass of the black hole $M_{\rm BH}(a_{\rm F})$. 
For a rough estimate we assume that all the $\psi$ particles within the horizon at $a_{\rm F}$ form a BH, with mass
\begin{equation}\label{eq:mass_BH}
M_{\rm BH}(a_{\rm F})=\frac{4\pi}{3} \frac{\rho_{\psi}(a_{\rm F})}{ H^{3}(a_{\rm F})}\simeq 
\frac{4\pi}{3\beta^2}\frac{M_P^2}{H_{\rm F}}\,.
\end{equation}
Here we have employed the value of $\Omega_\psi$ according to the scaling solution and
$H_{\rm F}\equiv H(a_{\rm F})$. Actual masses could be somewhat smaller than the estimate \eqref{eq:mass_BH} 
if the infalling mass covers only part of the horizon volume. This does not change orders of magnitude. 
Expressing $M_{\rm BH}$ in solar mass 
units $M_\odot$ and taking into account the value of the Hubble parameter at matter radiation equality, $H_{\rm eq}$, we get  
\begin{equation}\label{eq:mass_BH2}
 \frac{M_{\rm BH}(a_{\rm F})}{M_\odot}=\frac{c}{3\beta^2}\frac{H_{\rm eq}}{H_{\rm F}}\,, \hspace{8mm} 
 c\equiv
 \frac{M_{\rm eq}}{M_\odot}=2.7\times 10^{17}\,,
\end{equation}
with $M_{\rm eq}=2G H^{-1}_{\rm eq}$.
Thus BHs with ten solar masses form at an epoch with $H_{\rm F}\approx 10^{16}\beta^{-2} H_{\rm eq}$ or equivalently 
at a redshift $z_{\rm BH}\approx 10^{11}\vert \beta\vert^{-1}$, typically after nucleosynthesis, $z_{\rm NS}=10^{9}$. More 
massive BHs form even later. We conclude that 
our mechanism can produce BHs in the mass range observed by the LIGO and VIRGO collaborations only if the mass of the scalar 
field is below $H_{\rm F}\approx 2.4\times 10^{-12}\beta^{-2}$ eV, and if the initial conditions are such that $\Omega_\psi$
reaches the scaling solution only near nucleosynthesis. 

Let us next estimate the energy density in primordial BHs, $\rho_{\rm BH}$. Since BH formation proceeds 
very rapidly once the fluctuations $\delta_\psi$ become nonlinear, we may assume a complete conversion where all $\psi$ particles 
end in BHs. In this limit of instantaneous complete 
conversion one has $\rho_{\rm BH}(a_{\rm F})=\rho_\psi(a_{\rm F})$ and 
$\Omega_{\rm BH}(a_{\rm F})=\Omega_\psi(a_{\rm F})\approx1/(3\beta^2)$. If only a fraction $f$ of the $\psi$ particles ends in 
black holes our estimate of $\Omega_{\rm BH}$ has to be multiplied by $f$.
After formation $\Omega_{\rm BH}$ grows like nonrelativistic matter, such that at the end of the RD epoch one has 
\begin{equation} \label{eq:rel1}
\Omega_{\rm BH}(a_{\rm eq})=\frac{a_{\rm eq}}{a_{\rm F}}\Omega_{\rm BH}(a_{\rm F})=\frac{1}{3\beta^2}
\frac{a_{\rm eq}}{a_{\rm F}}\,.
\end{equation}
For $\Omega_{\rm BH}(a_{\rm eq})=1/2$, the BHs constitute all the dark matter in the Universe. An 
abundance $\Omega_{\rm BH}>1/2$  would lead to overclosure of the Universe putting bounds on the underlying models with light 
scalar fields. 

Combining Eqs.~\eqref{eq:mass_BH2} and \eqref{eq:rel1} we can express the typical mass of the 
produced BHs in terms of $\beta$ and $\Omega_{\rm BH}(a_{\rm eq})$,
\begin{equation}\label{eq:re;2}
\frac{M_{\rm BH}}{M_\odot}=\frac{c}{3\beta^2}\left(\frac{a_{\rm F}}{a_{\rm eq}}\right)^2 =
\frac{c}{27\beta^6\Omega^2_{\rm BH}(a_{\rm eq})}\,.
\end{equation}
The relations \eqref{eq:rel1} and \eqref{eq:re;2} fix both $M_{\rm BH}/M_\odot$ and $\Omega_{\rm BH}(a_{\rm eq})$ as functions 
of the parameters $\beta$ and $a_{\rm F}/a_{\rm eq}$. While $\beta$ is a model parameter, $a_{\rm F}/a_{\rm eq}$ depends on 
the initial conditions for $\rho_\psi$. For fixed $\beta$, smaller $\rho_{\psi,{\rm in}}$ leads to larger 
$a_{\rm F}/a_{\rm eq}$. 
In particular we may determine the value of $\beta_c$ for which dark matter is dominated by BHs
\begin{equation}\label{eq:Z}
\vert \beta_c\vert
=585\left(\frac{M_{\rm BH}}{M_\odot}\right)^{-1/6}=\left(\frac{2a_{\rm eq}}{3 a_{\rm F}}\right)^{1/2}\,.
\end{equation}
For $\beta^2\geq 2 a_{\rm eq}/(3 a_{\rm F}(\beta))$ an additional dark matter component is needed, while
models with $\beta^2\leq 2 a_{\rm eq}/(3 a_{\rm F}(\beta))$ are excluded.

We finally relate $a_{\rm F}/a_{\rm eq}$ to the initial conditions for $\Omega_\psi$ in different scenarios:

1. In our first 
scenario the $\psi$ particles are produced during the heating and entropy production after inflation. We will denote the end 
of the heating period by $a_{\rm ht}$. If $\Omega_\psi(a_{\rm ht})$ is of the order $1/(3\beta^2)$ the fluctuations become  
nonlinear at $a_{\rm F}/a_{\rm ht}=\exp({N_{\rm F}})$, or equivalently at  $a_{\rm F}/a_{\rm eq}=\exp(N_{\rm F}-N_{\rm eq})$
with $N_{\rm eq}$ the number of $e$-folds between the end of the heating period and matter-radiation equality. 
If we want to prevent the dimensionless coupling $g$ from being larger than one, $\beta^2$ is bounded by $M_P^2/m^2$. Since $m^2$ 
must be larger than  $H^2(a_{\rm ht})$, this puts a limit on the end of the heating period,
\begin{equation}\label{eq:AB}
N_{\rm eq}\lesssim \ln\left(\frac{M_P}{H_{\rm eq}}\right)^{2/5}+\frac{N_{\rm F}}{5}\simeq 50+\frac{N_{\rm F}}{5}\,.
\end{equation}
The critical value $\beta_c$ for BH dark matter is very large in this scenario,
\begin{equation}\label{eq:AC}
 \beta_c\approx \exp\left[(N_{\rm eq}-N_{\rm F})/2\right]\,,
\end{equation}
and according to Eq.~\eqref{eq:re;2} typical BH masses are tiny as compared to the solar mass.  For 
$\Omega_\psi(a_{\rm ht})$ larger than $1/(3\beta^2)$ the growth of fluctuations will be even faster, with smaller 
$N_{\rm F}$ and BH masses.

2. For a second scenario we assume $\Omega_\psi(a_{\rm ht})\ll 1/(3\beta^2)$, while the $\psi$ particles are already decoupled 
from radiation. In this case the onset of growth is delayed by the time $\Omega_\psi$ needs to grow to the order 
$1/(3\beta^2)$. This replaces $N_{\rm eq}\rightarrow N_{\rm eq}+\ln\left(3\beta^2\Omega_\psi(a_{\rm ht})\right)$
 in Eqs.~\eqref{eq:AB} and \eqref{eq:AC}. As can be seen from Fig.~\ref{fig:evol} the relation between $\rho_\psi$ before the 
 scaling solution and $\Omega_{\rm BH}(a_{\rm eq})$ is almost independent of $\beta$.
  
3. In a third scenario the $\psi$ particles are in thermal equilibrium at $a_{\rm ht}$. They decouple from radiation at some time $a_{\rm dc}$. At this time, their abundance can be strongly Boltzmann suppressed, such that $\Omega_\psi(a_{\rm dc})$ 
can be very small. Effectively, this replaces  in Eqs.~\eqref{eq:AB} and \eqref{eq:AC} 
$N_{\rm eq}\rightarrow N_{\rm eq}+\ln(3\beta^2\Omega_\psi(a_{\rm dc})a_{\rm ht}/a_{\rm dc})$.
The BH masses can now be substantially larger, and the LIGO and VIRGO range 
 can be realized for sufficiently small $\Omega_\psi(a_{\rm dc})$. 

\section{Conclusions}
In this paper we presented a new mechanism for black hole formation which does not rely on inflationary
physics. We argued that primordial black holes could be generated by
long-range interactions as those ubiquitously appearing in beyond the Standard Model scenarios or modified 
gravity/dark energy theories. For illustration purposes we considered a very simple scenario containing only 
a light scalar field coupled with some fermion field beyond the Standard Model. We showed that for sufficiently large couplings, the system enters
a scaling regime in which the fermion energy density \textit{tracks} the background component. During 
this regime the primordial fermion perturbations become significantly enhanced and can eventually collapse
into black holes. 
 
Can the produced black holes contribute to dark matter?   There are tight observational constraints
on the abundance of PBHs~\cite{ref:constraints}. According to Refs.~\cite{Carr:2016drx,Clesse:2017bsw}, 
there exists an open window for dark matter BHs around 1-1000 $M_{\odot}$ (see also
Refs.~\cite{Kuhnel:2017pwq,Carr:2017jsz}). Since this is also the interesting regime
to explain the gravitational wave detections, we will focus next on models for
which the BH mass distribution peaks at $M_{\rm max} \sim {\cal O} (M_{\odot})$. We 
emphasize, however, that this choice is made for illustration purposes only and that other 
ranges of masses can be easily accommodated by different choices of  $\beta$. The value of 
$M_{\rm max}$ depends indeed on the precise value of  this effective coupling. For BH dark 
matter [$\Omega_{\rm BH}(a_{\rm eq})=1/2$] it is well approximated by Eq.~\eqref{eq:Z},
\begin{equation}
M_{\rm max}\simeq \left(\frac{585}{\beta}\right)^6 M_\odot\,. \label{Mmaxfinal}
\end{equation}
A more refined analysis taking into account the entropy production between $a_{\rm F}$ and $a_{\rm eq}$, as well as the 
distribution of $\delta_\psi$ perturbations via a Press-Schechter-type formalism \cite{Press:1973iz},
changes the maximum mass \eqref{Mmaxfinal} by an ${\cal O}(1)$ multiplicative 
factor $\Delta g_s \left(p/(2+p)\right)^{1/p}$, with $\Delta g_s=(g_s(a_{\rm eq})/g_s(a_{\rm F}))^{1/2}$ and $g_s$ 
the number of entropic degrees of freedom. The mean and the standard deviation of the BH 
distribution are very close to $M_{\rm max}$, leading to a nearly monochromatic mass spectrum. 
To obtain masses in the range 1-1000 $M_{\odot}$, we must have  couplings in the 
range $185\lesssim \beta\lesssim 585 $. For the corresponding very small density 
parameter in the scaling regime, $\Omega_{\psi}=1/(3\beta^2)$, there are no constraints 
from nucleosynthesis. Black hole formation takes 
place at an epoch $a_{\rm F}=2 a_{\rm eq}/(3\beta^2)$, cf. Eq.~\eqref{eq:Z}, 
corresponding to temperatures of order  ${\cal O}({\rm MeV})$. The $\psi$-particles must be stable until this epoch. 

Our results are only rough order of magnitude estimates.  A more detailed  account of the
formation and evolution processes is needed.  On the one hand,  merging and accretion 
effects will tend to shift and broaden the BH mass distribution \cite{Chisholm:2005vm,Carr:2009jm}. 
On the other hand, extensions and modifications of the model with more than one heavy particles' species
may lead to nonmonocromatic spectra and modifications of the observational constraints \cite{Carr:2017jsz}. In 
view of the present uncertainties, a production of black holes in the range $1-1000\, M_\odot$ and 
constituting the whole dark matter component 
of the Universe seems possible. The necessary ingredients are a scalar field with mass 
smaller than $10^{-14}$ eV, heavy fields in a suitable abundance, and a mutual 
effective coupling of the order $\beta_c$.  The almost massless scalar field could be
the cosmon of dynamical dark energy \cite{Wetterich:1987fm,Wetterich:1994bg,Rubio:2017gty}, which has at all times a dynamical 
mass of the order $H$. 

\vspace{2mm}
\section*{Acknowledgements} We acknowledge support from the DFG through the project TRR33, ``The Dark Universe''.  

\bibliographystyle{plain}

\begin{thebibliography}{99}

\bibitem{Zeldovich1966}
Y.~B. Zel'dovich, I.~D. Novikov,
\href{http://adsabs.harvard.edu/abs/1966AZh....43..758Z}{Astron. Zh. 43, 758, (1966)}.

\bibitem{Hawking:1971ei}
  S.~Hawking,
  Mon.\ Not.\ Roy.\ Astron.\ Soc.\  {\bf 152} (1971) 75.
  
\bibitem{Carr:1975qj}
  B.~J.~Carr,
  Astrophys.\ J.\  {\bf 201} (1975) 1.
  
\bibitem{Chapline1975}
George F. Chapline, 
Nature 253, 251-252, 1975.

\bibitem{ref:LIGO}
  B.~P.~Abbott {\it et al.} [LIGO Scientific and Virgo Collaborations],
  Phys.\ Rev.\ Lett.\  {\bf 116} (2016) no.6,  061102;
  B.~P.~Abbott {\it et al.} [LIGO Scientific and Virgo Collaborations],
  Phys.\ Rev.\ Lett.\  {\bf 116} (2016) no.24,  241103;
  B.~P.~Abbott {\it et al.} [LIGO Scientific and VIRGO Collaborations],
  Phys.\ Rev.\ Lett.\  {\bf 116}, no. 20, 201301 (2016);
  S.~Bird, I.~Cholis, J.~B.~Muñoz, Y.~Ali-Haïmoud, M.~Kamionkowski, E.~D.~Kovetz, A.~Raccanelli and A.~G.~Riess,
  Phys.\ Rev.\ Lett.\  {\bf 116} (2016) no.20,  201301;
  S.~Blinnikov, A.~Dolgov, N.~K.~Porayko and K.~Postnov,
  JCAP {\bf 1611} (2016) no.11,  036.
     

\bibitem{Kawasaki:2012kn}
  M.~Kawasaki, A.~Kusenko and T.~T.~Yanagida,
  Phys.\ Lett.\ B {\bf 711} (2012) 1
  
\bibitem{Carr:2016drx}
  B.~Carr, F.~Kuhnel and M.~Sandstad,
  Phys.\ Rev.\ D {\bf 94} (2016) no.8,  083504
  
  
\bibitem{ref:inflection}
  J.~Garcia-Bellido, A.~D.~Linde and D.~Wands,
  Phys.\ Rev.\ D {\bf 54} (1996) 6040;
  J.~Yokoyama,
  Astron.\ Astrophys.\  {\bf 318} (1997) 673;
  T.~Nakamura, M.~Sasaki, T.~Tanaka and K.~S.~Thorne,
  Astrophys.\ J.\  {\bf 487} (1997) L139;
  M.~Drees and E.~Erfani,
  JCAP {\bf 1104} (2011) 005;
  M.~Kawasaki, A.~Kusenko, Y.~Tada and T.~T.~Yanagida,
  Phys.\ Rev.\ D {\bf 94} (2016) no.8,  083523;  
  J.~Garcia-Bellido and E.~Ruiz Morales,
  Phys.\ Dark Univ.\  {\bf 18} (2017) 47; 
      K.~Kannike, L.~Marzola, M.~Raidal and H.~Veerm\"ae,
  JCAP {\bf 1709} (2017) no.09,  020;
    C.~Germani and T.~Prokopec,
  Phys.\ Dark Univ.\  {\bf 18} (2017) 6;
    H.~Motohashi and W.~Hu,
  Phys.\ Rev.\ D {\bf 96} (2017) no.6,  063503.;
  G.~Ballesteros and M.~Taoso,
  Phys.\ Rev.\ D {\bf 97} (2018) no.2,  023501.

\bibitem{ref:bubble}
M.~Crawford and D.~N.~Schramm,
  Nature {\bf 298} (1982) 538;  
   S.~W.~Hawking, I.~G.~Moss and J.~M.~Stewart,
  Phys.\ Rev.\ D {\bf 26} (1982) 2681;   
  H.~Kodama, M.~Sasaki and K.~Sato,
  Prog.\ Theor.\ Phys.\  {\bf 68} (1982) 1979;  
   D.~La and P.~J.~Steinhardt,
  Phys.\ Lett.\ B {\bf 220} (1989) 375;   
  I.~G.~Moss,
  Phys.\ Rev.\ D {\bf 50} (1994) 676;  
  R.~V.~Konoplich, S.~G.~Rubin, A.~S.~Sakharov and M.~Y.~Khlopov,
  Phys.\ Atom.\ Nucl.\  {\bf 62} (1999) 1593;
  C.~J.~Hogan,
  Phys.\ Lett.\  {\bf 143B} (1984) 87;   
  S.~W.~Hawking,
  Phys.\ Lett.\ B {\bf 231} (1989) 237;   
  A.~Polnarev and R.~Zembowicz,
  Phys.\ Rev.\ D {\bf 43} (1991) 1106;   
  R.~R.~Caldwell and P.~Casper,
  Phys.\ Rev.\ D {\bf 53} (1996) 3002;  
   H.~B.~Cheng and X.~Z.~Li,
  Chin.\ Phys.\ Lett.\  {\bf 13} (1996) 317;   
  J.~H.~MacGibbon, R.~H.~Brandenberger and U.~F.~Wichoski,
  Phys.\ Rev.\ D {\bf 57} (1998) 2158;
  V.~A.~Berezin, V.~A.~Kuzmin and I.~I.~Tkachev,
  Phys.\ Lett.\  {\bf 120B} (1983) 91; 
  R.~R.~Caldwell, A.~Chamblin and G.~W.~Gibbons,
  Phys.\ Rev.\ D {\bf 53} (1996) 7103.
  
    
\bibitem{ref:Qballs}
    E.~Cotner and A.~Kusenko,
  Phys.\ Rev.\ Lett.\  {\bf 119} (2017) no.3,  031103;
    E.~Cotner and A.~Kusenko,
  Phys.\ Rev.\ D {\bf 96} (2017) no.10,  103002.

\bibitem{Magg:1980ut}
  M.~Magg and C.~Wetterich,
  Phys.\ Lett.\  {\bf 94B} (1980) 61.
  
\bibitem{Wetterich:1994bg}
  C.~Wetterich,
  Astron.\ Astrophys.\  {\bf 301} (1995) 321
 
\bibitem{Wetterich:2007kr}
  C.~Wetterich,
  Phys.\ Lett.\ B {\bf 655} (2007) 201
  
\bibitem{Amendola:2007yx}
  L.~Amendola, M.~Baldi and C.~Wetterich,
  Phys.\ Rev.\ D {\bf 78} (2008) 023015

\bibitem{Fardon:2003eh}
  R.~Fardon, A.~E.~Nelson and N.~Weiner,
  JCAP {\bf 0410} (2004) 005

\bibitem{Brookfield:2005bz}
  A.~W.~Brookfield, C.~van de Bruck, D.~F.~Mota and D.~Tocchini-Valentini,
  Phys.\ Rev.\ D {\bf 73} (2006) 083515
   Erratum: [Phys.\ Rev.\ D {\bf 76} (2007) 049901]

\bibitem{Casas:2016duf}
  S.~Casas, V.~Pettorino and C.~Wetterich,
  Phys.\ Rev.\ D {\bf 94} (2016) no.10,  103518
  
\bibitem{Amendola:1999er}
  L.~Amendola,
  Phys.\ Rev.\ D {\bf 62} (2000) 043511

\bibitem{Amendola:2003wa}
  L.~Amendola,
  Phys.\ Rev.\ D {\bf 69} (2004) 103524
  doi:10.1103/PhysRevD.69.103524


\bibitem{Ayaita:2012xm}
  Y.~Ayaita, M.~Weber and C.~Wetterich,
  Phys.\ Rev.\ D {\bf 87} (2013) no.4,  043519
  doi:10.1103/PhysRevD.87.043519

 
\bibitem{Ayaita:2011ay}
  Y.~Ayaita, M.~Weber and C.~Wetterich,
  Phys.\ Rev.\ D {\bf 85} (2012) 123010
  doi:10.1103/PhysRevD.85.123010
  
  
  
\bibitem{ref:constraints}  
  P.~Tisserand {\it et al.} [EROS-2 Collaboration],
  Astron.\ Astrophys.\  {\bf 469} (2007) 387; 
  M.~Ricotti, J.~P.~Ostriker and K.~J.~Mack,
  Astrophys.\ J.\  {\bf 680} (2008) 829;
  L~Wyrzykowski {\it et al.},
  Mon.\ Not.\ Roy.\ Astron.\ Soc.\  {\bf 458} (2016) no.3,  3012;
  E.~Mediavilla, J.~Jiménez-Vicente, J.~A.~Muñoz, H.~Vives-Arias and J.~Calderón-Infante,
  Astrophys.\ J.\  {\bf 836} (2017) no.2,  L18;
  V.~Poulin, P.~D.~Serpico, F.~Calore, S.~Clesse and K.~Kohri,
  Phys.\ Rev.\ D {\bf 96} (2017) no.8,  083524;
  Y.~Ali-Haïmoud and M.~Kamionkowski,
  Phys.\ Rev.\ D {\bf 95} (2017) no.4,  043534;
  T.~D.~Brandt,
  Astrophys.\ J.\  {\bf 824} (2016) no.2,  L31;
  A.~M.~Green,
  Phys.\ Rev.\ D {\bf 94} (2016) no.6,  063530;
  T.~S.~Li {\it et al.} [DES Collaboration],
  Astrophys.\ J.\  {\bf 838} (2017) no.1,  8.;
  Y.~Inoue and A.~Kusenko,
  JCAP {\bf 1710} (2017) no.10,  034.

\bibitem{Clesse:2017bsw}
  S.~Clesse and J.~García-Bellido,
  arXiv:1711.10458 [astro-ph.CO].
  
\bibitem{Kuhnel:2017pwq}
  F.~K\"uhnel and K.~Freese,
  Phys.\ Rev.\ D {\bf 95} (2017) no.8,  083508

\bibitem{Carr:2017jsz}
  B.~Carr, M.~Raidal, T.~Tenkanen, V.~Vaskonen and H.~Veerm\"ae,
  Phys.\ Rev.\ D {\bf 96} (2017) no.2,  023514
 
 
\bibitem{Press:1973iz}
  W.~H.~Press and P.~Schechter,
  Astrophys.\ J.\  {\bf 187} (1974) 425.
  
    
\bibitem{Chisholm:2005vm}
  J.~R.~Chisholm,
  Phys.\ Rev.\ D {\bf 73} (2006) 083504
  
\bibitem{Carr:2009jm}
  B.~J.~Carr, K.~Kohri, Y.~Sendouda and J.~Yokoyama,
  Phys.\ Rev.\ D {\bf 81} (2010) 104019
  
\bibitem{Wetterich:1987fm}
  C.~Wetterich,
  Nucl.\ Phys.\ B {\bf 302} (1988) 668
  
\bibitem{Rubio:2017gty}
  J.~Rubio and C.~Wetterich,
  Phys.\ Rev.\ D {\bf 96} (2017) no.6,  063509
  

\end{thebibliography}

\end{document}